 \documentclass[conference]{IEEEtran}

\usepackage{ifpdf}

\usepackage{cite}

\ifCLASSINFOpdf
  \usepackage[pdftex]{graphicx}
   \DeclareGraphicsExtensions{.pdf,.jpeg,.png}
\else
   \usepackage[dvips]{graphicx}
\fi
\usepackage[cmex10]{amsmath}
\usepackage{algorithmic}

\usepackage{array}

\usepackage{mdwmath}
\usepackage{mdwtab}

\usepackage{eqparbox}
\usepackage[caption=false]{subfig}

\usepackage{fixltx2e}

\usepackage{stfloats}

\usepackage{url}

\usepackage{graphicx}
\usepackage[autostyle]{csquotes}
\usepackage{mdwlist}
\newcommand{\bit}{\begin{itemize}}
\newcommand{\eit}{\end{itemize}}
\usepackage{amsmath}
\usepackage{amsfonts}
\usepackage{amssymb}

\begin{document}
%
\title{Genetic Sequence Matching Using D4M Big Data Approaches}

\author{\IEEEauthorblockN{Stephanie Dodson, Darrell O. Ricke, and Jeremy Kepner}
\IEEEauthorblockA{MIT Lincoln Laboratory, Lexington, MA, U.S.A\\
}}


%


\maketitle

\begin{abstract}
Recent technological advances in Next Generation Sequencing tools have led to increasing speeds of DNA sample collection, preparation, and sequencing.  One instrument  can produce  over 600 Gb of genetic sequence data in a single run.  This creates new opportunities to efficiently handle the increasing workload.  We propose a new method of fast genetic sequence analysis using the Dynamic Distributed Dimensional Data Model (D4M) -- an associative array environment for MATLAB developed at MIT Lincoln Laboratory.  Based on mathematical and statistical properties, the method leverages big data techniques and the implementation of an Apache Acculumo database to accelerate computations one-hundred fold over other methods. Comparisons of the D4M method with the current gold-standard for sequence analysis, BLAST, show the two are comparable in the alignments they find.  This paper will present an overview of the D4M genetic sequence algorithm and statistical comparisons with BLAST.
\end{abstract}


%
\IEEEpeerreviewmaketitle
\let\thefootnote\relax\footnote{This work is sponsored by the Assistant Secretary of Defense for Research and Engineering under Air Force Contract \#FA8721-05-C-0002.  Opinions, interpretations, recommendations and conclusions are those of the authors and are not necessarily endorsed by the United States Government.}

\section{Introduction}
New technologies are producing an ever-increasing volume of sequence data, but the inadequacy of current tools puts restrictions on large-scale analysis.  At over 3 billion base pairs (bp) long, the human genome naturally falls into the category of Big Data, and techniques need to be developed to efficiently analyze and uncover novel features. Applications include sequencing individual genomes, cancer genomes, inherited diseases, infectious diseases, metagenomics, and zoonotic diseases.

In 2003, the cost of sequencing the first human genome was \$3 billion. The cost for sequencing has declined steadily since then and is projected to drop to \$100 in several years. The dramatic decrease in the cost of obtaining genetic sequences has resulted in an explosion of data with a range of applications, including early identification and detection of infectious organisms from human samples.

DNA sequences are highly redundant among organisms. For example, Homo sapiens share about 70\% of genes with the zebrafish \cite{zebrafish}. Despite the similarities, the discrepancies create unique sequences that act as fingerprints for organisms. The magnitude of sequence data makes correctly identifying organisms based on segments of genetic code a complex computational problem.

\begin{figure*}[!t] 
\centerline{\subfloat[]{\includegraphics[width =2.7in]{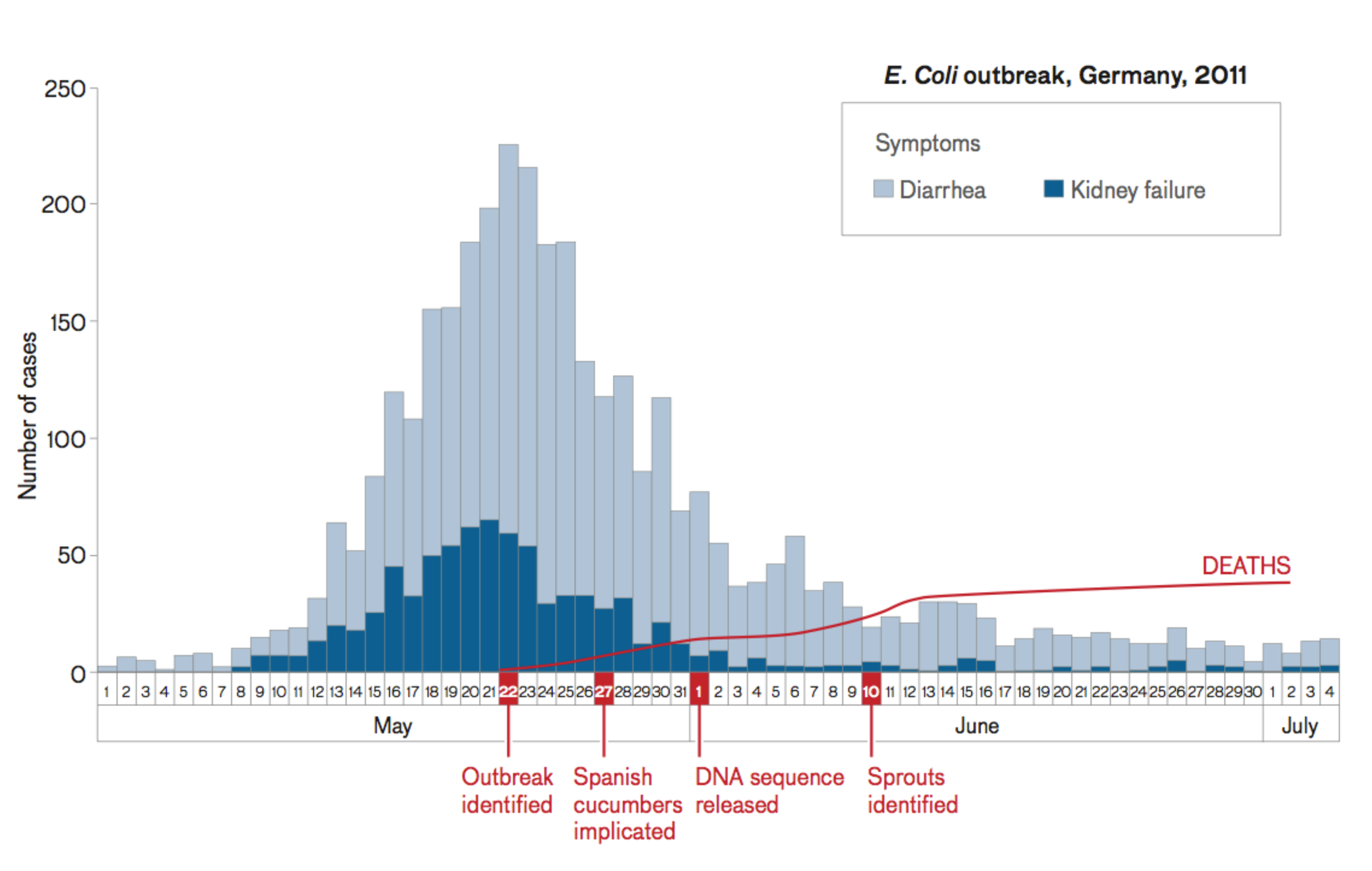} 
\label{ecolia}} 
\hfil 
\subfloat[]{\includegraphics[width=3in]{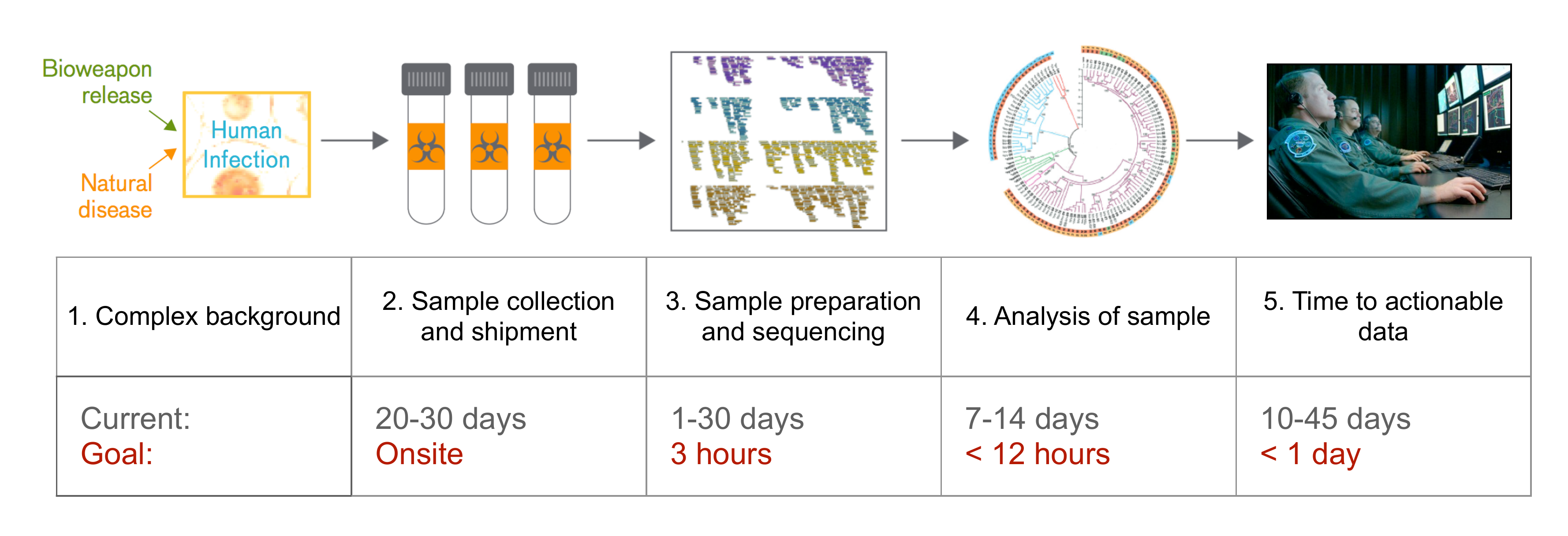}
 \label{ecolib}}} 
\caption{Example disease outbreak (a) and processing pipeline (b). In the May to July 2011 virulent \emph{E. coli} outbreak in Germany, the identification of the \emph{E. coli} source was too late to have substantial impact on illnesses. Improved computing and algorithms can play a significant role in reducing the current time of 10 to 45 days to less than 1 day. Photo source: http://defensetech.org/2012/09/05/the-dangers-of-the-pentagons-cloud. } 
\label{timeline} 
\end{figure*}

Given one segment of DNA, current technologies can quickly determine what organism it likely belongs to.  However, the speed rapidly diminishes as the number of segments increases. The time to identify all organisms present in a human sample (blood or oral swab), can be up to 45 days (Figure \ref{ecolib}) with major bottlenecks being (1) sample shipment to sequencing centers and (2) sequence analysis.

Consider the specific case of an \emph{E. coli} outbreak that took place in Germany in the summer of 2011 (Figure \ref{ecolia}).  In this instance, while genetic sequencing resolved how the virulence of this isolate was different with the insertion of a bacterial virus carrying a toxic gene from previously characterized isolates, it arrived too late to significantly impact the number of deaths.

Figure \ref{ecolib} shows that there are a number of places in the “Current” time frames for the sequencing process where improved algorithms and computation could have a significant impact. Ultimately, with the appropriate combination of technologies, it should be possible to shorten the timeline for identification from weeks to less than one day. A shortened timeline could significantly reduce the number of deaths from such an outbreak.

\subsection{BLAST}
First developed in 1990, the Basic Local Alignment Search Tool (BLAST) is the current gold-standard method used by biologists for genetic sequence searches. Many versions are available on the National Center for Biotechnology Information (NCBI) website to match nucleotide or protein sequences \cite{blast_online}.  In general, BLAST is a modification of the Smith -- Waterman algorithm \cite{smith_waterman} and works by locating short \enquote{seeds} where the reference and unknown samples have identical seed sequences \cite{statistics}\cite{blast_nature}. Seeds are expanded outwards, and with each added base pair, probabilities and scores are updated to signify how likely it is the two sequences are a match. Base pairs are added until the probabilities and scores reach threshold values \cite{blast_nature}\cite{blast}.

\begin{figure*}[!b]
\centering
\includegraphics[width=7in]{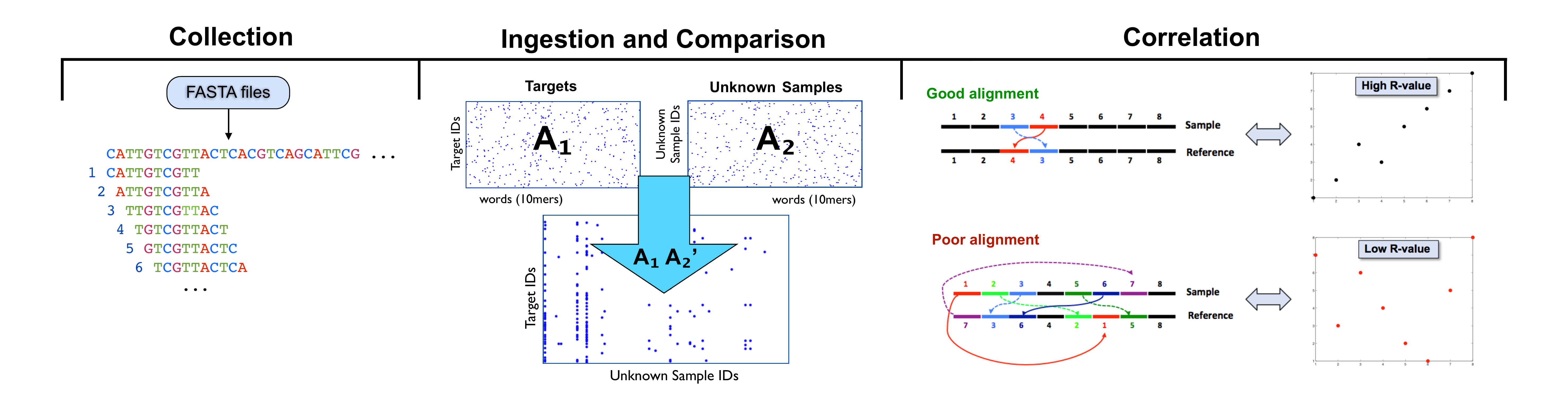}
\caption{Pipeline for analysis of DNA sample using D4M technologies. Data is received from the sequencers in FASTA format, and parsed into 10-mers.  The row, column, and value triples are ingested into D4M associative arrays, and matrix multiplication finds the common words between sample and reference sequences.  Matches are tested for good alignment using the linear R-value correlation. }
\label{steps}
\end{figure*}

The score most highly used by biologists to distinguish true matches is the expect value, or E-value.  The E-value uses distributions of random sequences and sequence lengths to measure the expected number of matches that will occur by chance.  Similar to a statistical P-value, low E-values represent matches \cite{statistics}\cite{blast_nature}\cite{blast}.  E-value thresholds depend on the dataset, with typical values ranging from 1E-6 to 1E-30. Despite the relation to the P-value, E-value calculations are complex and do not have a straightforward meaning.  Additionally, the underlying distributions are known to break down in alignments with shorter sequences or numerous repeats (low-complexity sequences) \cite{statistics}.

Direct use of BLAST to compare a 600 Gb collection with a comparably sized reference set requires months on a 10,000 core system. The Dynamic Distributed Dimensional Data Model (D4M) developed at Lincoln Laboratory, has been used to accelerate DNA sequence comparison.

\section{Computational Methods}

\subsection{D4M Schema}
D4M is an innovation in computer programming that combines the advantages of five processing technologies: triple store databases, associative arrays, distributed arrays, sparse linear algebra, and fuzzy algebra. Triple store databases are a key enabling technology for handling massive amounts of data and are used by many large Internet companies (e.g., Google Big Table) \cite{bigTable}. Triple stores are highly scalable and run on commodity computers, but lack interfaces to support rapid development of the mathematical algorithms used by signal processing experts. D4M provides a parallel linear algebraic interface to triple stores. Using D4M, developers can create composable analytics with significantly less effort than if they used traditional approaches. The central mathematical concept of D4M is the associative array that combines spreadsheets, triple stores, and sparse linear algebra. Associative arrays are group theoretic constructs that use fuzzy algebra to extend linear algebra to words and strings \cite{D4Minfo}.

\begin{figure*}[!t] 
\centerline{\subfloat[]{\includegraphics[width =2.7in]{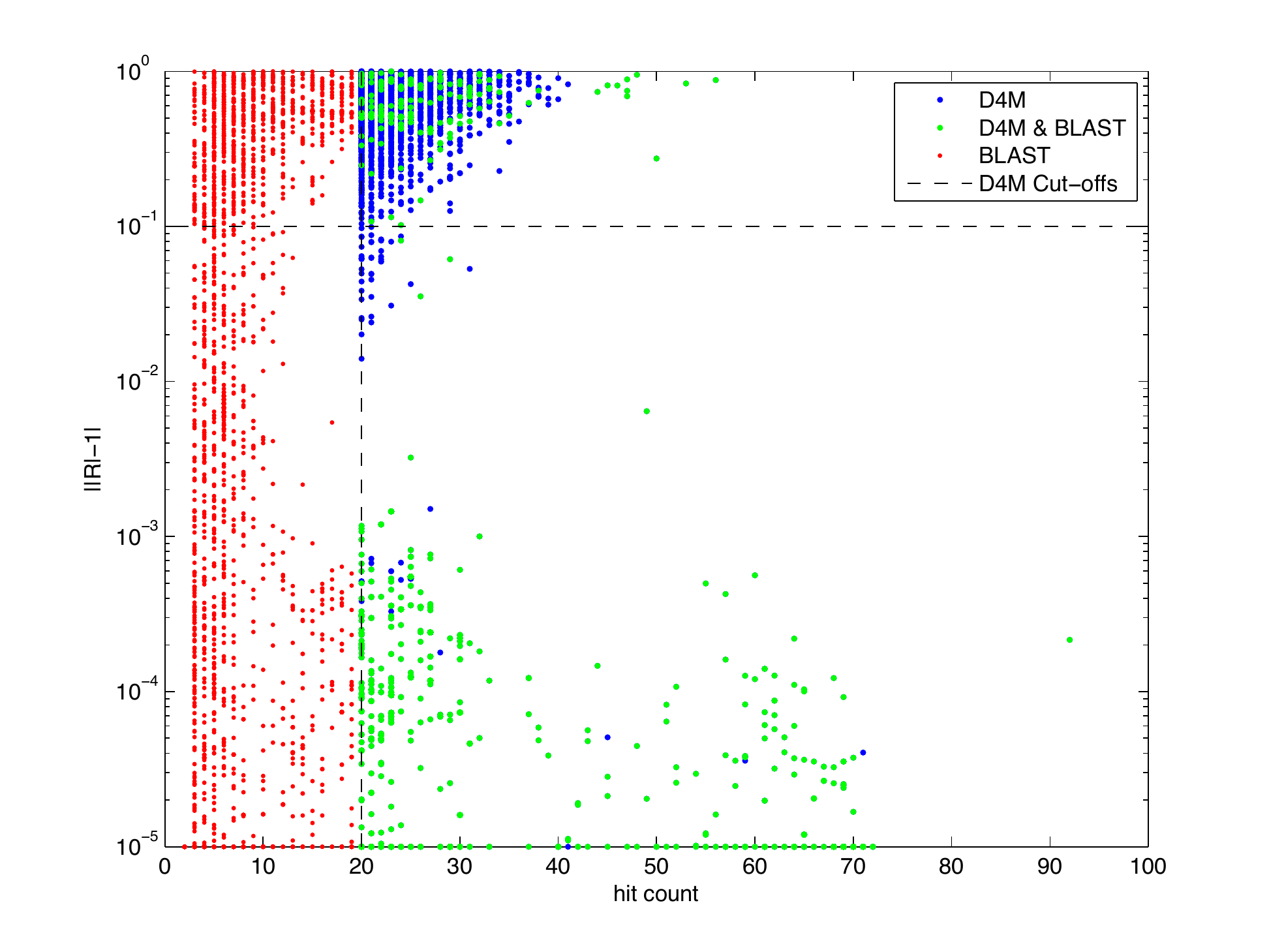} 
\label{figure:logR_hc}} 
\hfil 
\subfloat[]{\includegraphics[width=2.7in]{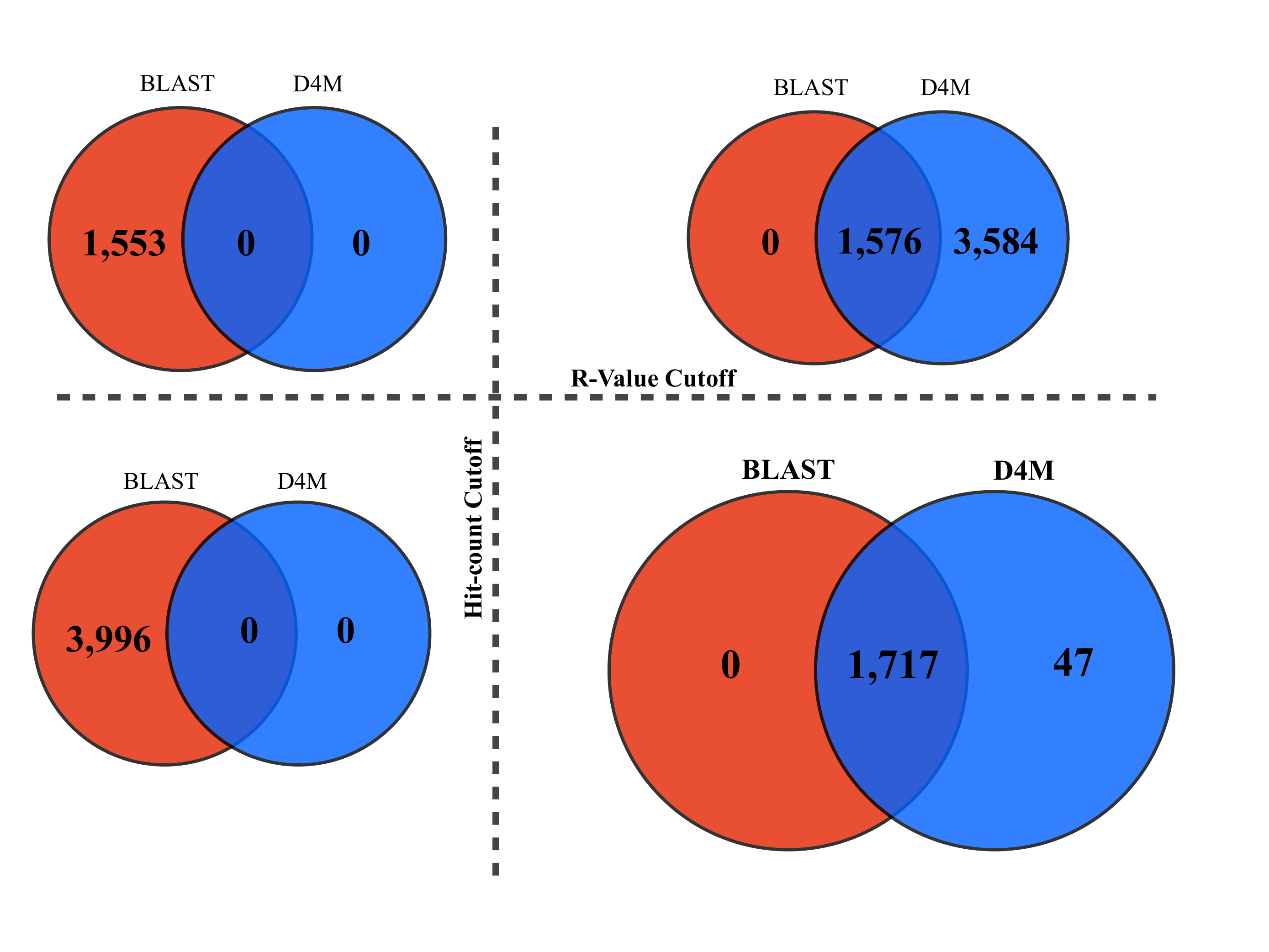}
 \label{figure:statistics}}} 
\hfil
\subfloat[]{\includegraphics[width=6.5in]{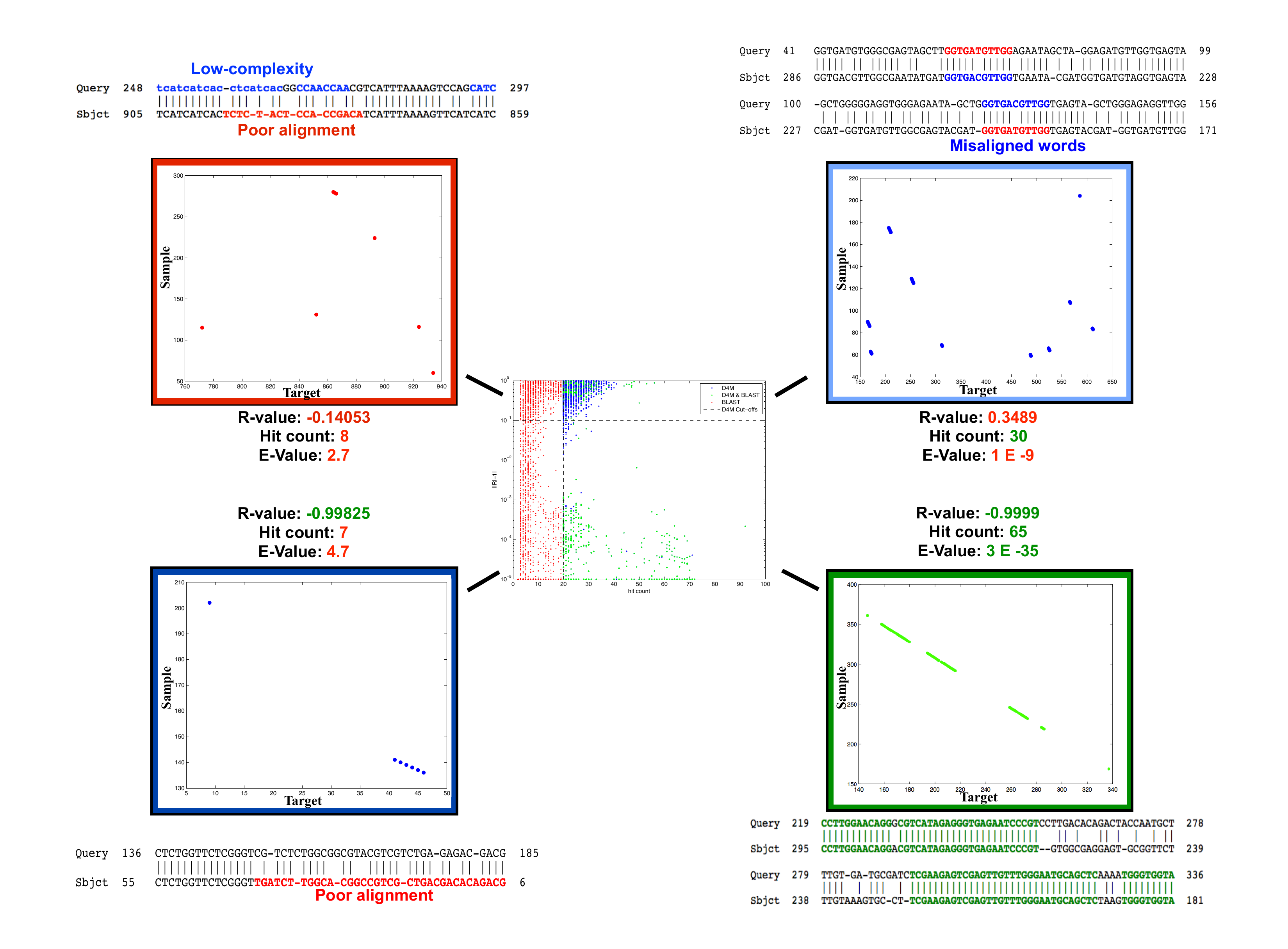}
\label{figure:E7examples} } 
\caption{{Quality of matches based on number of common words and R-values. \textbf{(a)} The vertical axis displays the modified R-value ($||R|-1|$) in a log scale, and the horizontal is the hit count. Due to selection requirements (dashed lines), the strong matches lie in the bottom right quadrant of the graph. \textbf{(b)} The arrangement of the figure mirrors (a) and shows the number of matches found by D4M and BLAST in each region.  \textbf{(c)} Representatives matches from each quadrant show the relationship between R-values, hit counts, and BLAST alignments. Reliable BLAST matches have an E-score below 10$^{-30}$. Low hit counts and R-values are the result of large stretches of poor alignment and low-complexity repeats (top-left). Clockwise to the top-right, hit counts improve with increased complexity, but misaligned words result in a scattering of uncorrelated points.  The words highlighted in blue and red are misaligned. At the bottom-left, higher R-values are created by segments of good alignment, but low hit counts remain due to large portions of poor alignment. Strong matches (bottom-right) are the result of long stretches of good alignment.  The straight green bands in the graph correspond to the correctly aligned highlighted lengths of DNA.} 
\label{alignments_all} }
\end{figure*}

Associative arrays provide an intuitive mechanism for representing and manipulating triples of data and are a natural way to interface with the new class of high performance NoSQL triple
store databases (e.g., Google Big Table, Apache Accumulo, Apache HBase, NetFlix Cassandra, Amazon Dynamo) \cite{D4Maccumulo}. 

Because the results of all queries and D4M functions are associative arrays, all D4M expressions are composable and can be directly used in linear algebraic calculations. The composability of associative arrays stems from the ability to define fundamental mathematical operations whose results are also associative arrays \cite{D4Minfo}. Given two associative arrays A and B, the results of all the following operations will also be associative arrays:

\indent A + B \hfill A - B  \hfill A \& B \hfill A$|$B \hfill A*B \hfill

D4M provides tools that enable the algorithm developer to implement a sequence alignment algorithm on par with BLAST in just a few lines of code. The direct interface to high performance triple store databases allows new database sequence alignment techniques to be explored quickly.

\subsection{D4M Algorithm}
The presented algorithm is designed to couple linear algebra approaches implemented in D4M with well-known statistical properties to simplify and accelerate current methods of genetic sequence analysis. The analysis pipeline can be broken into four key steps: collection, ingestion, comparison, and correlation (Figure \ref{steps}). 

 In \emph{collection}, unknown sample data is received from a sequencer in FASTA format, and parsed into a suitable form for D4M. DNA sequences are split into k-mers of 10 bases in length (words). Uniquely identifiable metadata is attached to the words and positions are stored for later use.  Low complexity words (composed of only 1 or 2 DNA bases) are dropped. Very long sequences are segmented into groups of 1,000 k-mers to reduce non-specific k-mer matches.  

\begin{figure*}[!t] 
\centerline{\subfloat[]{\includegraphics[width =3in]{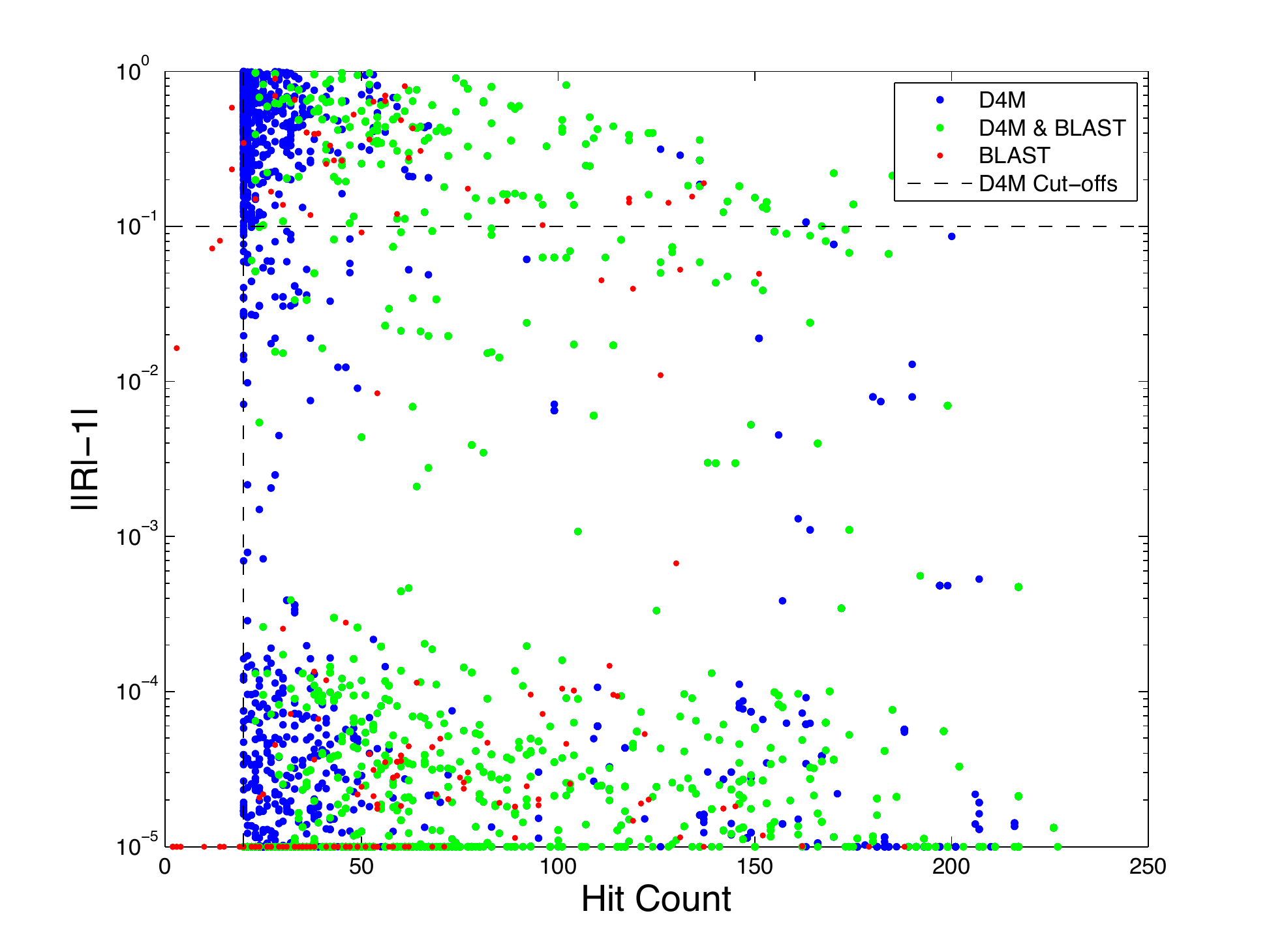} 
\label{logR_hcT7}} 
\hfil 
\subfloat[]{\includegraphics[width=3in]{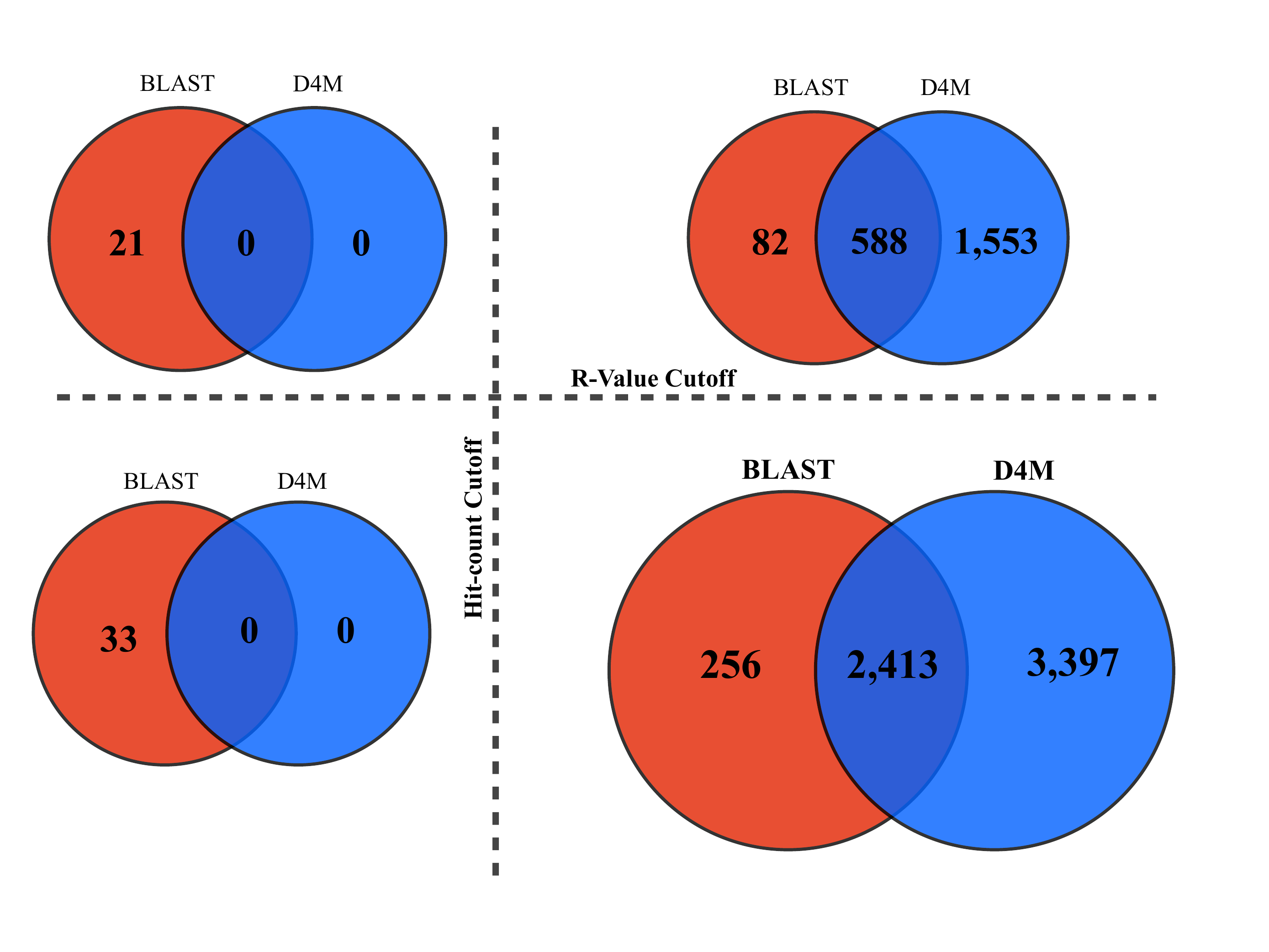}
 \label{statisticsT7}}} 
\caption{Distribution of modified R-values and words in common between two sequences for Dataset 2. \textbf{(a)} The vertical axis displays $||R|-1|$ in a log scale, and the horizontal is the hit count. Dashed lines are D4M selection requirements, and true matches lie at the bottom right quadrant of the graph. Shown are all matches after top hit count selection and prior to final R-value cut. \textbf{(b)} The arrangement of the figure mirrors (a) and shows the number of matches found by D4M and BLAST in each region.} 
\label{hitcountsT7} 
\end{figure*}

The \emph{ingestion} step loads the sequence identifiers, words, and positions into D4M  associative arrays by creating unique rows for every identifier and columns for each word. The triple store architecture of associative arrays effortlessly handles the ingestion and organization of the data. During the process, redundant k-mers are removed from each 1,000 bp segment, and only the first occurrence of words are saved. The length and the four possible bases gives a total of 4$^{10}$ (just over one-million) possible words, naturally leading to sparse matrices and operations of sparse linear algebra.  Additionally, the segmentation into groups of 1,000 ensures sparse vectors/matrices with less than 1 in 1,000 values used for each sequence segment. A similar procedure is followed for all known reference data, and is also ingested into a matrix. 

Sequence similarity is computed in the \emph{comparison} step.  Using the k-mers as vector indices allows a vector cross-product value of two sequences to approximate a pair-wise alignment of the sequences. Likewise, a matrix multiplication allows the comparison of multiple sequences to multiple sequences in a single mathematical operation. For each unknown sequence, only strong matches (those with greater than 20 words in common) are stored for further analysis. Computations are accelerated by the sparseness of the matrices.

The redundant nature of DNA allows two unrelated sequences to have numerous words in common. Noise is removed by assuring the words of two matching sequences fall in the same order.  The \emph{correlation} step makes use of the 10-mer positions to check the alignments.  When the positions in reference and unknown sequences are plotted against each other, true alignments are linearly correlated with an absolute correlation coefficient (R-value) close to one. Matches with over 20 words in common and absolute R-values greater than 0.9 are considered strong. After these initial constraints are applied, additional tests may be used for organism identification and with large datasets.

\section{Results}

The algorithm was tested using two generated datasets (Datasets 1 and 2) \cite{datasets}.  The smaller Dataset 1 (72,877 genetic sequences) was first compared to a fungal dataset and used to examine the selection criteria.  Results were compared with those found by running BLAST. Dataset 2 (323,028 sequences) was formed from a human sample and spiked with \emph{in silico} bacteria organisms using FastqSim \cite{anna}.  Bacterial results from Dataset 2 were again compared to BLAST and used to test for correct organism identification.  In both comparisons, the reference sets were compiled from RNA present in GenBank. It is important to note that the reference sequences are unique to the taxonomic gene level.  Therefore, each gene of an organism is represented by a unique sequence and metadata. 

\begin{figure*}[!t] 
\centerline{
\subfloat[]{\includegraphics[width=3in]{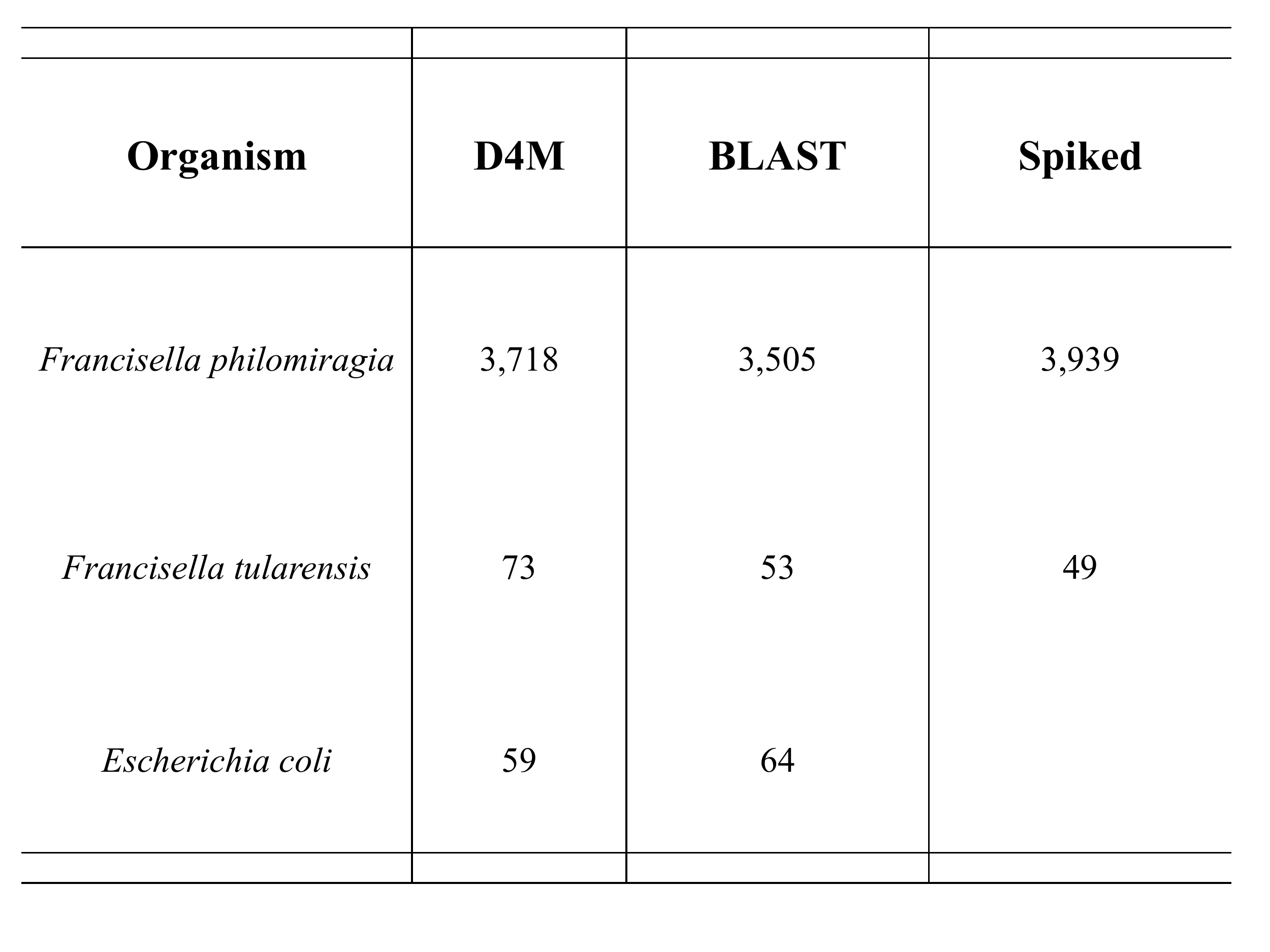}
 \label{table:noMatches}}  
\hfil 
\subfloat[]{\includegraphics[width =3in]{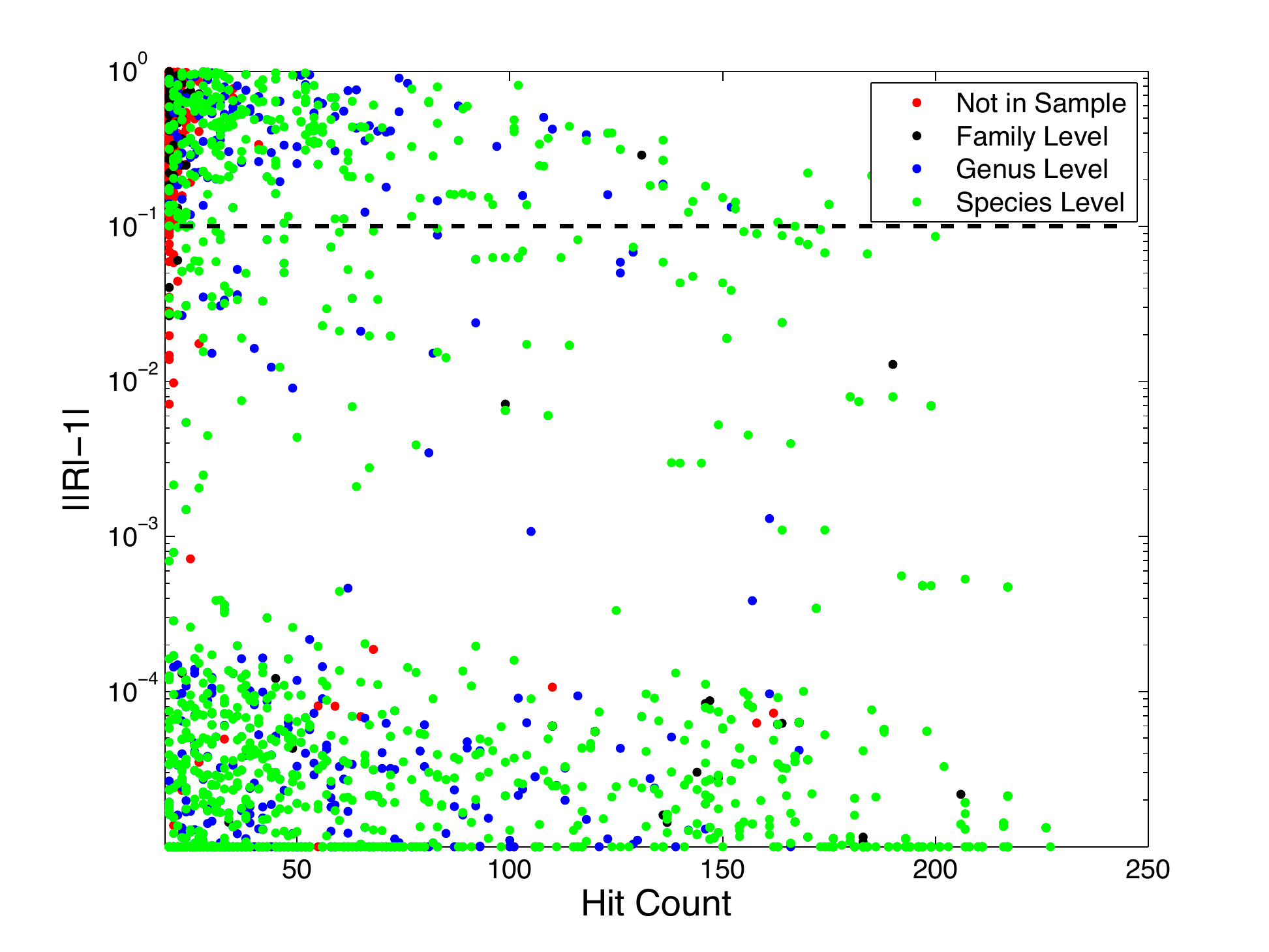} 
\label{sub:genusMatches}} }
\caption{Results of Dataset 2 shown in tabular (a) and graphical (b) forms. \textbf{(a)} Number of sequences D4M and BLAST found matching to the \emph{in silico} spiked data as well as a background organism (\emph{E. coli}) detected by both algorithms.  \textbf{(b)} Matches are highlighted based on correct identification of genus, species, strain, or false positives. Higher hit counts and R-values tend to correctly identify the genus and species, with lower values matching to artifacts.} 
\label{genusMatches} 
\end{figure*}

\subsection{Dataset 1}

Dataset 1 was compared to the fungal RNA dataset and, BLAST was run using BLASTN and a MIT Lincoln Laboratory developed Java BlastParser program. D4M found 6,924 total matches, while BLAST discovered 8,842. Examination demonstrates the quality of the matches varies based on the hit counts and correlation coefficients. To separate the minute differences in linear correlation, the R-values were modified to Log($||R|-1|$). Because of this choice, the strong R-values with absolute value close to one have a modified value near zero.  Figure \ref{figure:logR_hc} displays the modified R-values plotted versus the hit counts.  The dashed lines show the threshold values of 20 words in common and a R-value greater than $|0.9|$. Strong matches lie in the bottom right of the graph. The unusually large void between 10$^{-2}$ and 10$^{-1}$ on the vertical axis (R-values of 0.9 to 0.99) serves as a clear distinction between regions of signal and noise for the D4M and BLAST data.

Together, the hit count and R-value thresholds greatly reduce the background noise. Figure \ref{figure:statistics} shows a full distribution of the number of matches satisfying the numerical requirements. Almost 63\% of the total BLAST finds have hit counts less than 20, and about 18\%  fall below both D4M thresholds. Before the correlation selection, D4M identifies 5,160 background matches, of which, BLAST finds 1,576. After all restrictions, D4M and BLAST both identify 1,717 matches.

 Representative correlations from each region demonstrate the selection quality of the algorithm (Figure \ref{figure:E7examples}).  Also shown are the BLAST alignments, in which the top strand is the unknown sequence and the bottom is the matching reference. Vertical lines indicate an exact base pair alignment. Dashes in the sequences represent a gap that was added by BLAST to improve the arrangement. The alignments include the initial seeds and the expansion until threshold values were reached. 

The results demonstrate how hit counts below twenty are a result of poor alignments and low complexity repeats; these matches have few regions with at least a 10 bp overlap. Strong alignments as seen in the bottom-right of Figure \ref{figure:E7examples} are comprised of long, well aligned stretches.  Segments of poor alignment still exist, but they are proportionally less, and offset the sample and reference sequences by equal amounts. It is worth noting that the BLAST E-values of the four examples coincide with the D4M results. The results of Dataset 1 present comparable findings to BLAST run with default parameters.

\subsection{Dataset 2}
Additional filters were used in the analysis of the larger Dataset 2. As in Dataset 1, alignments were first required to have at least 20 words in common. Additionally, for each unknown sequence, the R-value was only computed for matches within 10\% of the maximum hit count. For example, unknown sequence A might have 22 words in common with reference B, 46 with reference C, and 50 with reference D. R-values were calculated for the matches with references C and D since the hit counts are within 10\% of 50, the maximum value. Similar to Dataset 1, absolute R-values were thresholded at 0.9, but were also required to be within 1\% of the maximum for each unknown sequence. The additional percentage threshold values were chosen to identify matches of similar strength and reduce the number of computations.

Again, results were compared with BLAST, this time run with default parameters. Comparisons of BLAST and D4M findings before R-value filters are displayed in Figure \ref{logR_hcT7}. The stricter BLAST conditions eliminate many of the false positives with low hit counts and R-values as seen in Dataset 1. Before the R-value restrictions, D4M finds significantly more alignments than BLAST (Figure \ref{statisticsT7}).  These numbers are reduced with R-value filters, but during organism identification steps, the majority of the additional points mapped to the correct species (discussed below). Notice the majority of BLAST findings missed by D4M lie in the lower hit count regime, all of which emerge from the second hit count filter (within 10\% of maximum). 

After applying all filters, each sample matched either to one or multiple references. Unique matches were labeled as that reference.  In the case of multiple alignments, the taxonomies were compared and the sample was classified as the lowest common taxonomic level. For example, if unknown sequence A maps equally to references B and C, both of which are different species within the same genus, sequence A is classified as the common genus. The number of sequences matching to each family, genus, and species is tallied to give the final results.

In Figure \ref{table:noMatches}, D4M and BLAST results are numerically compared with the spiked organisms. The numbers indicate how many sequences were classified as species.  Both D4M and BLAST correctly identified the species \emph{F. philomiragia} and \emph{F. tularensis}, with numbers close to the truth data.  It is important to note that \emph{F. philomiragia} and \emph{F. tularensis} are very closely related. Studies show \emph{F. philomiragia} and \emph{F. tularensis} to have between 98.5\% and 99.9\% identity \cite{Ribosomal_analysis} which accounts for the slight difference in numbers of matching sequences. D4M and BLAST identified nearly the same amount of \emph{E. coli} presence. Interestingly, \emph{E. coli} was not an \emph{in silica} spiked organism, and is instead a background organism (present in the human sample) detected by both.

As previously noted, the numbers in Figure \ref{statisticsT7} show D4M identified significantly more matches than BLAST. The D4M data in Figure \ref{logR_hcT7} was color-coded based on the taxonomy of matches. Results are presented in Figure \ref{sub:genusMatches} and reveal the majority of points with high hit counts and R-values are matching to the truth data. Again, at this stage, no R-value filters have been applied, but the results clearly indicate how hit counts and R-values are appropriate selection parameters to correctly and efficiently identify organisms present in a sample.


\subsection{Computational Acceleration with Apache Accumulo}

In the analysis described, parallel processing using pMatlab \cite{pMatlab} was heavily relied upon to increase computation speeds.  The implementation of an Apache Accumulo database additionally accelerated the already rapid sparse linear algebra computations and comparison processes, but was not used to the full advantage.

\begin{figure}[!t]
\centering
\includegraphics[width=3.2in]{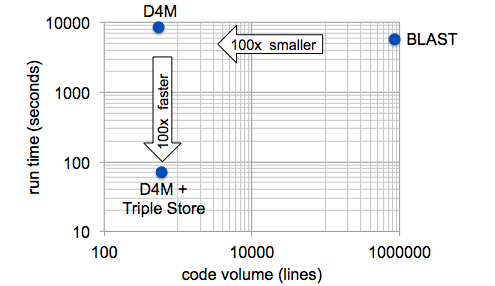}
\caption{Sequence-alignment implementations. The D4M implementation requires 100x less code than BLAST. D4M + Triple Store (Accumulo) reduces run time by 100x compared to BLAST. }
\label{speed}
\end{figure}

As shown in \cite{Taming_Big_Data}, the triple store database can be used to identify the most common 10-mers. The least popular words are then selected and used in comparisons, as these hold the most power to uniquely identify the sequences. Preliminary results show subsampling greatly reduces the number of direct comparisons, and increases the speed 100x. Figure \ref{speed} shows the relative performance and software size of sequence alignment implemented using BLAST, D4M alone, and D4M with triple store Accumulo. Future developments will merge the results discussed here with the subsampling acceleration techniques of the database.

\section{Conclusion}

With Matlab and D4M techniques, the described algorithm is implemented in less than 1,000 lines of code. That gives a 100x improvement over BLAST, and on comparable hardware the performance level is within a factor of 2. The precise code allows for straight forward debugging and comprehension. Results shown here with Datasets 1 and 2 demonstrate that D4M findings are comparable to BLAST and possibly more accurate.

The next steps are to integrate the Apache Accumulo capabilities and optimize the selection parameters over several known datasets. Additionally, the capabilities will be ported to the SciDB database. The benefit of using D4M in this application is that it can significantly reduce programming time, increase performance, and simplify the current complex sequence matching algorithms.

\section*{Acknowledgment}

The authors are indebted to the following individuals for their technical contributions to this work: Chansup Byun, Ashley Conard, Dylan Hutchinson, and Anna Shcherbina.



%

\end{document}